\documentclass{lhep}
\journal{LHEP}
\vol{xx}
\jyear{2024}
\pages{xxx}

\usepackage{amssymb}
\usepackage{graphicx}
\usepackage{times}
\usepackage{xspace}
\usepackage{array}
\usepackage{bm}
\usepackage{color}
\usepackage{enumerate}
\newcommand{\BE}{\begin{equation}}
\newcommand{\EE}{\end{equation}}

\begin{document}

\title{A new 700 GeV scalar in the LHC data?}

\author{Maurizio Consoli,\auno{1} George Rupp,\auno{2}}
\address{$^1$INFN, Sezione di Catania, I-95123 Catania, Italy}
\address{$^2$CeFEMA and CFTP, Instituto
Superior T\'{e}cnico, Universidade de Lisboa, P-1049-001 Lisboa, Portugal}

\begin{abstract}
As an alternative to the metastability of the electroweak vacuum,
resulting from perturbative calculations, one can consider a
non-perturbative effective potential which, as at the beginning of
the Standard Model, is restricted to the pure $\Phi^4$ sector yet
consistent with the known analytical and numerical studies. In this
approach, where the electroweak vacuum is now the lowest-energy
state, besides the resonance of mass $m_h=$ 125~GeV defined by the
quadratic shape of the potential at its minimum, the Higgs field
should exhibit a second resonance with mass
$(M_H)^{\rm Theor}=690\,(30)$~GeV associated with the zero-point
energy determining the potential depth. In spite of its large mass,
this resonance would couple to longitudinal $W$s with the same typical
strength as the low-mass state at 125~GeV and represent a relatively
narrow resonance, mainly produced at LHC by gluon-gluon fusion. In this
Letter, we review LHC data suggesting a new resonance of mass
$(M_H)^{\rm EXP} \sim 682\,(10)$~GeV, with a statistical significance
that is far from negligible.
\end{abstract}

\maketitle


\section{Introduction}
\label{intro}

The discovery \cite{discovery1,discovery2} of a narrow scalar
resonance with mass $m_h\!=\! 125$~GeV and the phenomenological
consistency with the expectations for the Higgs boson have confirmed
spontaneous symmetry breaking (SSB) through the Higgs field as a
fundamental ingredient of current particle physics. Yet, in our
opinion, the present perturbative description of symmetry breaking
is not entirely satisfactory. Indeed, once we start to look farther
than the Fermi scale, such a calculation predicts that the effective
potential of the Standard Model (SM) should have a new minimum, well
beyond the Planck scale and being much deeper than the electroweak (EW)
vacuum; see e.g.\ \cite{branchina,gabrielli}. While it is reassuring
that the most accurate calculation \cite{degrassi} gives a tunnelling
time that is larger than the age of the universe, still the idea of
a metastable vacuum raises several questions: the role of
gravitational effects, the mechanism explaining why the theory
remains trapped in our EW vacuum, \ldots. This would require a
cosmological perspective and to control the properties of matter in
the extreme conditions of the early universe.

As an alternative, one can consider
\cite{Consoli:2020nwb,symmetry,memorial,rupp} a non-perturbative
effective potential which, as at the beginning of the SM,
is restricted to the pure $\Phi^4$ sector but is also
consistent with the known analytical and numerical studies. In this
approach, where the EW vacuum is now the lowest-energy
state, besides the resonance of mass $m_h=125$~GeV defined by the
quadratic shape of the potential at its minimum, the Higgs field
should exhibit a second resonance with a mass
$(M_H)^{\rm Theor} = 690\,(30)$~GeV, associated with the zero-point
energy (ZPE) determining the potential depth. This large $M_H$
stabilises the potential, because including the ZPEs of all known
gauge and fermion fields would now represent just a small radiative
correction.\footnote{By
subtracting quadratic divergences or using dimensional
regularisation, the logarithmically divergent terms in the ZPEs of the
various fields are proportional to the fourth power of the mass.
Thus, in units of the pure scalar term, one finds
$(6 M^4_w+3 M^4_Z)/M^4_H \lesssim 0.002$ and $12m^4_t/M^4_H\lesssim 0.05$.}
Like in the early days of the SM, one could thus adopt the
perspective of explaining SSB within the pure scalar sector.
Checking this picture then requires to observe the second
resonance and its phenomenology.

In this respect, the hypothetical $H$ is not like a regular Higgs boson of
700~GeV, because it would couple to longitudinal $W$s with the same
typical strength as the low-mass state at 125~GeV
\cite{Consoli:2020nwb,symmetry,memorial,rupp}. In fact, it is
$m_h\!=\!125$~GeV (and not $M_H\!\sim\!700$~GeV) which fixes the quadratic
shape of the potential and the interaction with the Goldstone bosons. Thus,
the large conventional widths $\Gamma(H\!\to\!ZZ+WW)\sim G_F M^3_H$
would be suppressed by the small ratio $(m_h/M_H)^2\!\sim\! 0.032$,
leading to the estimates
$\Gamma(H\!\to\!ZZ)\sim M_H/(700\,{\rm GeV}) \times (1.6\:{\rm GeV})$ and
$\Gamma(H\to WW)\sim M_H/(700\,{\rm GeV}) \times (3.3\:{\rm GeV})$. As such,
the heavy $H$ should be a relatively narrow resonance of total width
$\Gamma(H\!\to\!{\rm all})=25\div 35$~GeV, decaying predominantly to
$t\bar t$ quark pairs, with a branching ratio of about 70$\div80\,$\%.
Due to its small coupling to longitudinal $W$s, $H$ production through
vector-boson fusion (VBF) would also be negligible as compared to
gluon-gluon fusion (ggF), which has a typical cross section
$\sigma^{\rm ggF} (pp\!\to\!H) \sim 1100\,(170)$~fb \cite{widths,yellow},
depending on the QCD and $H$-mass uncertainties.

After this brief review, in this Letter we will present LHC data
suggesting a new resonance in the expected mass and width range,
with a non-negligible statistical significance. Partial evidence was
already presented in Refs.~\cite{signals,ichep,CFF,universe}, but in Sec.~2
we will consider a larger data set and also refine the analysis of
some final states. Finally, Sec.~3 will contain a summary and our
conclusions.

\section{Experimental signals from LHC}

Having a definite prediction $(M_H)^{\rm Theor} = 690\,(30)$~GeV, we
will look for signals in the expected region of invariant mass
600$\div 800$~GeV, so that local deviations from the background
should not be downgraded by the``look elsewhere'' effect. In this
search, one should also keep in mind that, with the present energy
and luminosity of LHC, the second resonance is too heavy to be seen
unambiguously in all possible channels. (Recall the $h(125)$
discovery, which initially was producing no signals in the
important $b\bar b$ and $\tau^+ \tau^-$ channels).

Now, with an expected large branching ratio
$B(H \to t\bar t)=(70\div 80)\,$\%, the natural starting point would be
the $t \bar t$ channel. However, in the relevant region
$m(t\bar t)=620\div820$~GeV, CMS measurements \cite{CMS_top} give a
background cross section $\sigma(pp\to t \bar t)= 107 \pm 7.6$~pb, which
is about 100 times larger than the expected signal
$\sigma(pp\to H \to t \bar t) \lesssim 1$~pb. Therefore, in
Refs.~\cite{signals,ichep,CFF,universe} the analysis was focused on
available channels with relatively smaller background, namely:
\begin{enumerate}[i)]
\item
ATLAS ggF-like 4-lepton events;
\item
ATLAS high-mass inclusive $\gamma\gamma$ events;
\item
ATLAS and CMS $(b\bar b+\gamma\gamma)$ events;
\item
CMS $\gamma\gamma$ pairs produced in $pp$ diffractive
scattering.
\end{enumerate}

\subsection{The ATLAS  ggF-like 4-lepton events}

To start with, we will review ATLAS work
\cite{ATLAS2,atlas4lHEPData} dedicated to the charged 4-lepton
channel and to the search for a heavy scalar resonance $H$ decaying
through the chain $H\to ZZ \to 4l$. This is important because, for
this particular channel, there is a precise prediction
characteristic of our picture. Indeed, for a relatively narrow
resonance, the resonant peak cross section
$\sigma_R (pp\to H\to 4l)$ can be well approximated as
\BE
\sigma_R (pp\to H\to 4l) \sim
\sigma (pp\to H) \, B( H \to ZZ) \, 4 B^2(Z \to l^+l^-) \, ,
\label{exp3}
\EE
with $4 B^2(Z \to l^+l^-)\sim 0.0045$. Thus, in our
case, where
$\Gamma(H \to ZZ)\sim M_H/(700\,{\rm GeV})\times(1.6\:{\rm GeV})$ and
$\sigma (pp\to H)\sim \sigma^{\rm ggF} (pp\to H) \sim 1100\,(170)$~fb,
by introducing the reduced width $\gamma_H=\Gamma(H\to{\rm all})/M_H$,
we find the sharp correlation
\BE
[\gamma_H \times \sigma_R (pp\to H\to 4l)^{\rm Theor}\sim(0.011\pm0.002)\,
{\rm fb} \; ,
\label{exp34}
\EE 
to be compared with the ATLAS data.

In the ATLAS search, the 4-lepton events were divided into ggF-like
and VBF-like events. By expecting our second resonance to be
produced through gluon-gluon fusion, we have considered the ggF-like
category. The only sample that satisfies the two requirements, viz.\
a) being homogeneous from the point of view of the selection and b)
having sufficient statistics, is the so called ggF-low category,
which contains a mixture of all three final states. Since for an
invariant mass around 700~GeV, the energy resolution of these events
varies considerably,\footnote{The resolution varies from about
12~GeV for $4e$ events to 19~GeV for $2e2\mu$, up to 24~GeV for
$4\mu$.} it is natural to adopt a large-bin visualisation to avoid
spurious fluctuations between adjacent bins. The numbers of events
are reported in Table~\ref{leptontable}
\cite{ATLAS2,atlas4lHEPData}.

\begin{table}
\tbl{For luminosity 139 fb$^{-1}$, we give the \mbox{ATLAS} ggF-low
events $\rm{N}_{\rm EXP}(E)$ and the estimated background
$\rm{N}_{\rm B}(E)$ \cite{ATLAS2,atlas4lHEPData} for invariant mass
$m_{4l}=E=530\div 830$~GeV. To avoid spurious migrations between
neighbouring bins, we group the data into bins of 60~GeV,
corresponding to the 10 bins of 30~GeV from $545\,(15)$~GeV to
$815\,(15)$~GeV; see Ref.~\cite{atlas4lHEPData}.
\label{leptontable}}
{\begin{tabular}{cccc}
$\rm E$[GeV] & $\rm{N}_{\rm EXP}(E)$  & $\rm{N}_{\rm B}(E)$ &
$\rm{N}_{\rm EXP}(E) - \rm{N}_{\rm B}(E)$ \\
\hline \hline
560(30) & 38$\pm 6.16$ & 32.0 & $ 6.00 \pm 6.16$  \\
\hline
620(30) & 25$\pm 5.00$ & 20.0 & $5.00 \pm 5.00$  \\
\hline
680(30) & 26$\pm 5.10$ & 13.04 & $12.96 \pm 5.10$ \\
\hline
740(30) & 3$\pm 1.73$ & 8.71& $-5.71 \pm 1.73$  \\
\hline
800(30) & 7$\pm 2.64$ & 5.97 & $1.03 \pm 2.64$  \\
\hline
\end{tabular}}
\end{table}
Now, subtracting the background from the observed events gives a
considerable excess, at about 680~GeV, and then a sizable defect,
around 740~GeV. A simple explanation for these two simultaneous
features would be the existence of a resonance of mass $M_H\sim700$~GeV
which, above the Breit-Wigner peak, produces the
characteristic negative-interference pattern proportional to
$(M^2_H-s)$. To check this idea, we have exploited the basic model
where the $ZZ$ pairs, each subsequently decaying into a charged
$l^+l^-$ pair, are produced by various mechanisms at the parton
level. For $E\equiv m(4l)$, this produces a smooth distribution of
background events $N_b(E)$, proportional to a background cross
section $\sigma_b(E)$, which can interfere with a resonance of mass
$M_H$ and total decay width $\Gamma_H$. The total cross section
$\sigma_T(E)$ can then be expressed as \cite{memorial}
\begin{eqnarray}
\sigma_T(E) & \!\!\!\!= & \!\!\!\!\sigma_b(E) + \frac{2(M^2_H -s)\,
\Gamma_H M_H}{(s-M^2_H)^2 +(\Gamma_H M_H)^2}\,\sqrt{\sigma_R\sigma_b(E)}\;+
\nonumber \\
&& \frac{(\Gamma_H M_H)^2}{(s-M^2_H)^2+(\Gamma_H M_H)^2} \, \sigma_R \; ,
\label{sigmat}
\end{eqnarray}
where we have introduced the resonant peak cross section
$\sigma_R$ at $s=M^2_H$.
\begin{figure}[ht]
\centering
\includegraphics[trim = 0mm 0mm 0mm 0mm,clip,width=7.0cm]
{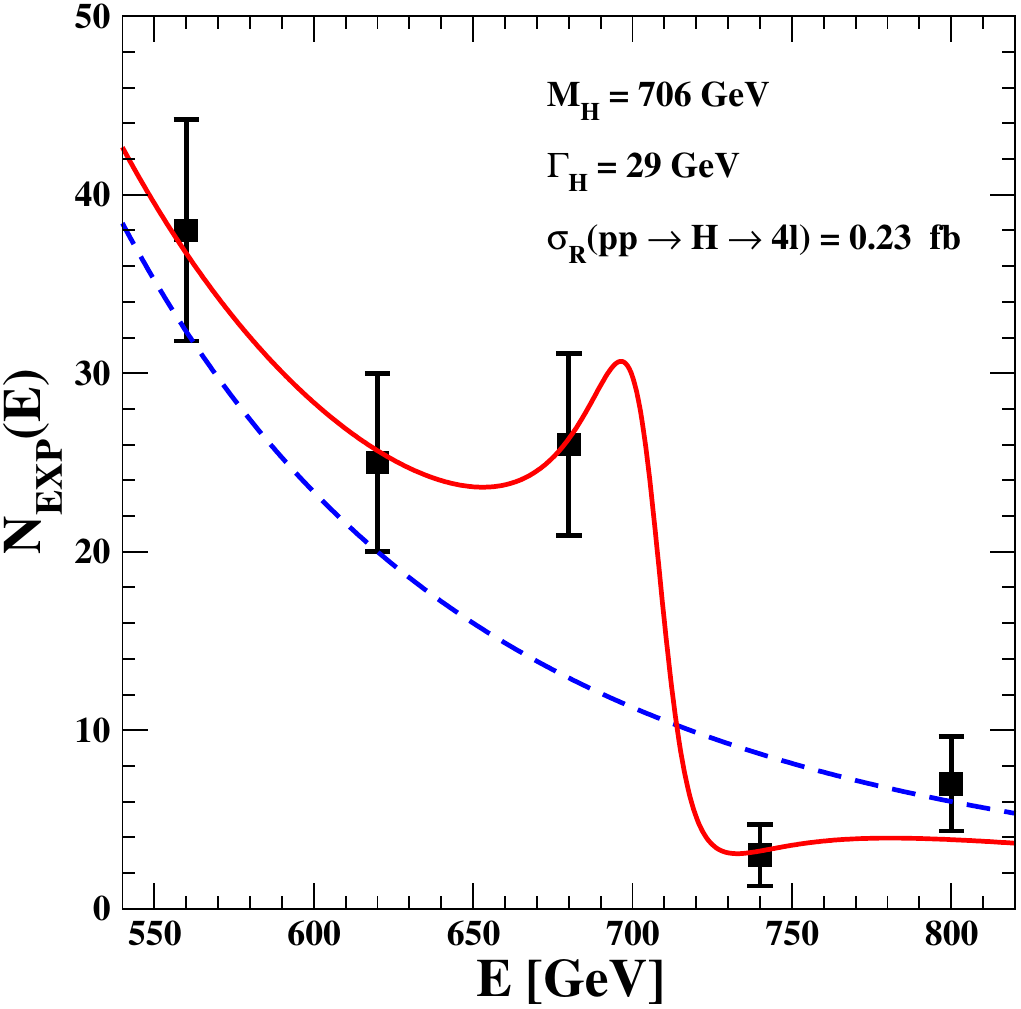}
\caption{The values $\rm{N}_{\rm EXP}(E)$
in Table~\ref{leptontable} vs.\ the corresponding $\rm{N}_{\rm TH}(E)$
in Eq.~(\ref{NTH}) (the full red curve). The resonance
parameters are $M_H= 706$~GeV, $\gamma_H=$0.041, $\sigma_R= 0.23$~fb
and the ATLAS background (dashed blue curve) is approximated as
$N_b(E)= A\times ({\rm 710~GeV}/E) ^{\nu}$, with $A= 10.55$ and
$\nu= 4.72$.}
\label{4lepton}
\end{figure}
Then, by simple redefinitions, the theoretical number of events can
be expressed as
\BE
N_{TH}(E) \; = \; N_b(E) + \frac{P^2 + 2 P \,
x(E) \, \sqrt{N_b(E)} }{\gamma^2_H + x^2(E)} \; ,
\label{NTH}
\EE
where $x(E)\!=\!(M^2_H -E^2)/M^2_H$, $P\!\equiv\!\gamma_H\sqrt{N_R}$, and
$N_R=\sigma_R \, {\cal A } \, 139$~fb$^{-1}$ denotes the extra
events at the resonance peak for an acceptance ${\cal A }$. The
data in Table~\ref{leptontable} were fitted with Eq.~(\ref{NTH}) in
\cite{CFF,universe}. The results are: $M_H= 706\,(25)$~GeV, $P= 0.14 \pm 0.07$
and $\gamma_H= 0.041\pm 0.029$, corresponding to a total width
$\Gamma_H= 29\pm 20$~GeV. One thus obtains  $N_R\sim 12^{+15}_{-9}$ and for 
acceptance $\langle{\cal A }\rangle\sim 0.38$, by averaging the two extremes
0.30 and 0.46 for the ggF-like category of events \cite{ATLAS2}, $\sigma_R\sim
0.23^{+0.28}_{-0.17}$~fb. A graphical comparison is shown in
Fig.~\ref{4lepton}.

The quality of the fit is good, but error bars are large and a test
of our picture is not very stringent. Still, with our
$\Gamma(H\to ZZ)\sim 1.6$~GeV, for $M_H \sim  700$~GeV, and the central
fitted $\langle\Gamma_H\rangle=$ 29~GeV, we find a branching ratio
$B(H\to ZZ)\sim 0.055$ which, for the theoretical value
$\sigma^{\rm ggF} (pp\to H)\sim 923$~fb of Ref.~\cite{yellow}
(at $M_H= 700$~GeV) would imply a theoretical peak cross section
$(\sigma_R)^{\rm Theor}=923\times0.055\times0.0045\sim0.23$~fb,
coinciding with the central value from our fit. Also, from the central
values $\langle\sigma_R\rangle= 0.23$~fb and
$\langle\gamma_H\rangle=0.041$, we find
$\langle\sigma_R\rangle \, \langle\gamma_H\rangle\sim 0.0093$~fb,
consistent with Eq.~(\ref{exp34}).

\begin{figure}[ht]
\centering
\includegraphics[trim = 0mm 0mm 0mm 0mm,clip,width=7.0cm]
{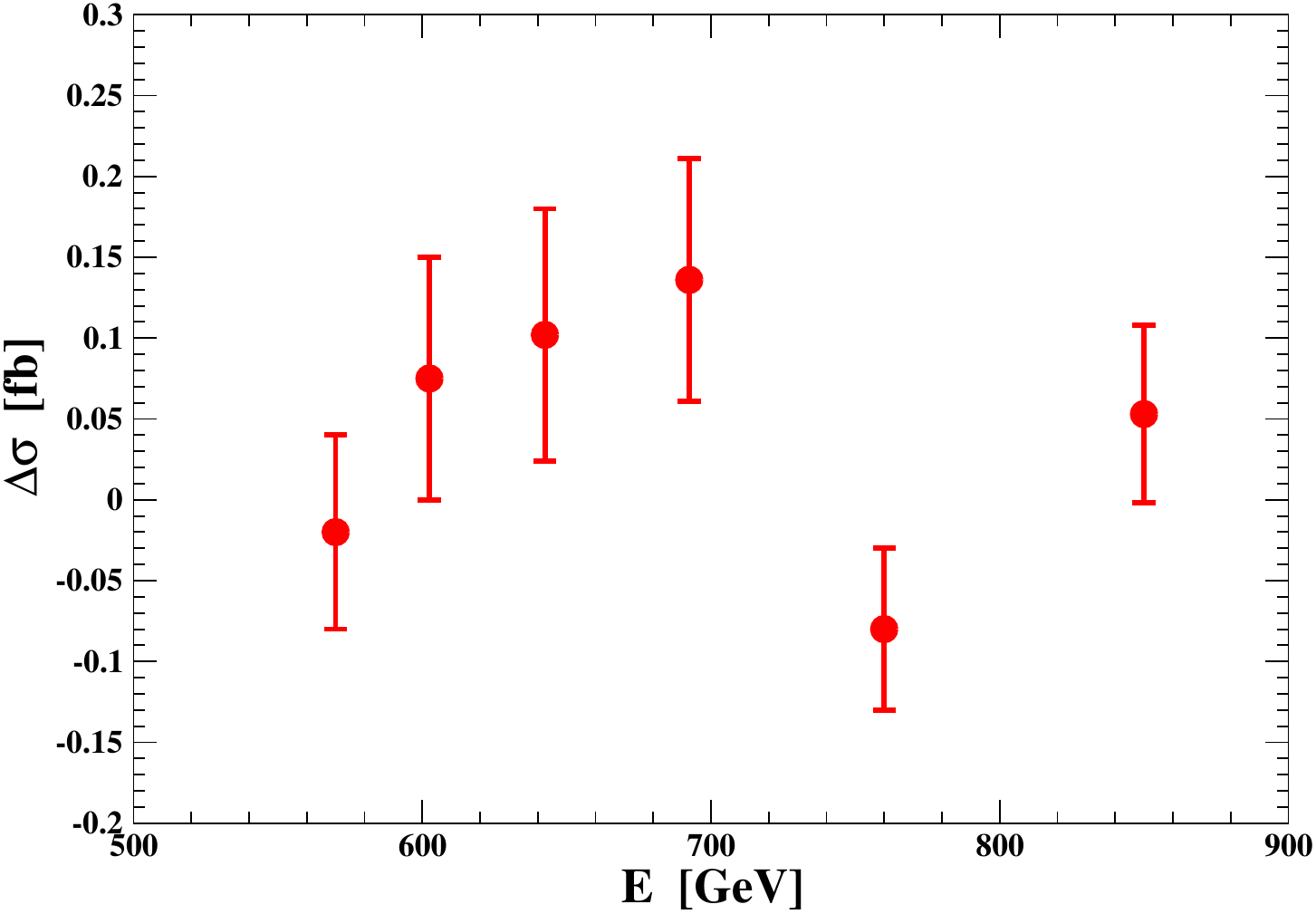}
\caption{The quantity $\Delta\sigma=(\sigma_{\rm EXP}-{\sigma}_{\rm B})$
given for each bin in the last column of Table~\ref{leptonxsection}.}
\label{Deltasigma}
\end{figure}
\begin{table}
\tbl{The observed ATLAS cross section
\cite{atlasnew,atlasnewdatafile} and estimated background for a 4-lepton
invariant mass $m(4l)\equiv E$ from 555 to 900~GeV. These values are
obtained by multiplying the bin size with the average differential cross
sections $\langle(d\sigma/dE)\rangle$, reported for each bin in the
companion HEPData file \cite{atlasnewdatafile}. The
full background cross section $\sigma_B$ contains a dominant
contribution from $ q\bar q \to 4l$ events.
\label{leptonxsection}}
{\begin{tabular}{cccc} Bin [GeV]& ${\sigma}_{\rm EXP} $~[fb]  &
$\sigma_{\rm B}$~[fb] &
$(\sigma_{\rm EXP} - {\sigma}_{\rm B} $)~[fb]  \\
\hline \hline
555--585 & 0.252 $\pm 0.056$ & $0.272 \pm 0.023$ & $-0.020 \pm 0.060$  \\
\hline
585--620 & $0.344\pm 0.070$ & $ 0.259 \pm 0.021 $ & $+0.085 \pm 0.075$ \\
\hline
620--665 & $0.356 \pm 0.075$ & $ 0.254 \pm 0.023$ & $+0.102\pm 0.078$  \\
\hline
665--720 & $0.350 \pm 0.073$ &$ 0.214 \pm 0.019$  &$+0.136 \pm 0.075$ \\
\hline
720--800 & $0.126 \pm 0.047$ &$ 0.206 \pm 0.018$ & $ -0.080 \pm 0.050$ \\
\hline
800--900 & $0.205\pm 0.052$ &$ 0.152 \pm 0.017$  & $+0.053 \pm 0.055$  \\
\hline
\end{tabular}}
\end{table}
After this first comparison from Refs.~\cite{CFF,universe}, we will now
consider the other ATLAS paper Ref.~\cite{atlasnew,atlasnewdatafile}
for the differential 4-lepton cross section $d \sigma/d E$, with
$E=m(4l)$. From Fig.~5 of this other Ref.~\cite{atlasnew}, one finds
the same excess-defect sequence as in Table~\ref{leptontable} and
additional support for a new resonance. The corresponding data are
given in Table~\ref{leptonxsection} and in Fig.~\ref{Deltasigma}.
Besides, by comparing with Ref.~\cite{atlasnew,atlasnewdatafile}, we
can sharpen our analysis. In fact, the ggF-low events considered
above contain a large contribution from $ q\bar q \to 4l$
processes. Although the initial state is $pp$ in all cases, our $H$
resonance would mainly be produced through gluon-gluon fusion, so
that, strictly speaking, the interference should only be computed
with the non-resonant $gg \to 4l$ background. Obtaining this
refinement is now possible, because, in
Ref.~\cite{atlasnew,atlasnewdatafile},  the individual contributions to
the background are reported separately. Denoting by
${\sigma}^{\rm gg}_{\rm B} $ the pure non-resonant $gg\to 4l$
background cross section, we can thus consider a corresponding
experimental cross section ${\hat \sigma}_{\rm EXP}$ after
subtracting out preliminarily the ``non-ggF' background:
\BE
{\hat \sigma}_{\rm EXP} = \sigma_{\rm EXP} -
({\sigma}_{\rm B} - {\sigma}^{\rm gg}_{\rm B} ) \; .
\label{sigmahat}
\EE
The corresponding values and background are given in
Table~\ref{redefinedxsection}.

\begin{table}
\tbl{For each energy bin, we give the ATLAS experimental cross
section ${\hat \sigma}_{\rm EXP}$ (cf.\ Eq.~(\ref{sigmahat})). The two
cross sections ${\sigma}_{\rm EXP} $ and ${\sigma}_{\rm B} $ are
listed in Table~\ref{leptonxsection}. The other background cross
section ${\sigma}^{\rm gg}_{\rm B} $ takes only into account the
non-resonant $gg\to 4l$ process and is computed by multiplying
the bin size with the average differential cross section
$(d {\sigma}^{\rm gg}_{\rm B}/dE)$ in each bin; see
Ref.~\cite{atlasnewdatafile}.
\label{redefinedxsection}}
{\begin{tabular}{cccc}
Bin [GeV] & ${\hat \sigma}_{\rm EXP} $~[fb] & ${\sigma}^{\rm gg}_{\rm B}$~[fb]
& $ (\hat\sigma_{\rm EXP}$ -${\sigma}^{\rm gg}_{\rm B}$)~[fb]  \\
\hline \hline
555-585& 0.003 $\pm 0.060$ & $0.023 \pm 0.004$  & $ -0.020 \pm 0.060 $  \\
\hline
585-620 & $0.105 \pm 0.073$ &$ 0.020 \pm 0.003$ & $+0.085  \pm 0.075$  \\
\hline
620-665&  $0.121 \pm 0.078$ &$ 0.019 \pm 0.003$ & $+0.102  \pm 0.078$  \\
\hline
665-720 & $0.152\pm 0.075$ &$ 0.016 \pm 0.003$ & $+0.136 \pm 0.075$  \\
\hline
720-800& $-0.067\pm 0.050$ &$ 0.013 \pm 0.002$ & $-0.080  \pm 0.050$  \\
\hline
800-900 & $0.062\pm 0.055$ &$ 0.009 \pm 0.002$ & $+0.053  \pm 0.055$  \\
\hline
\end{tabular}}
\end{table}

\begin{table}
\tbl{
Comparing the experimental cross section in Table~\ref{redefinedxsection}
with the theoretical Eq.~(\ref{sigmat}) for the optimal set of parameters
$M_H= 677$~GeV, $\Gamma_H= 21$~GeV, $\sigma_R=0.40$~fb.
\label{comparexsection}}
{\begin{tabular}{cccc}
Bin [GeV] & ${\hat \sigma}_{\rm EXP} $~[fb] & $ \sigma_T $~[fb] & $\chi^2$ \\
\hline \hline
555-585& 0.003 $\pm 0.060$ & 0.048 & 0.56  \\
\hline
585-620& $0.105 \pm 0.073$ & 0.056 &  0.45  \\
\hline
620-665& $0.121 \pm 0.078$ & 0.123 &   0.00  \\
\hline
665-720& $0.152\pm 0.075$ & 0.152 &   0.00  \\
\hline
720-800& $-0.067\pm 0.050$ & 0.002 &   1.90  \\
\hline
800-900 & $0.062\pm 0.055$ & 0.004 &   1.11 \\
\hline
\end{tabular}}
\end{table}

We then compare the resulting experimental ${\hat \sigma}_{\rm EXP}$ with
the theoretical $\sigma_T$ in Eq.~(\ref{sigmat}), after
the identification $\sigma_b= {\sigma}^{\rm gg}_{\rm B}$. By
parametrising the ATLAS differential background
$(d\sigma^{\rm gg}_{\rm B}/dE) \sim A \times ({\rm 710~GeV}/E)^{\nu}$,
with $A\sim (2.42 \pm 0.18)\times 10^{-4}$~fb/GeV and $\nu\sim 5.24 \pm 0.45$,
and integrating the various contributions to Eq.~(\ref{sigmat})
within each energy bin, a fit to the data gives
$M_H=677^{+30}_{-14}$~GeV, $\Gamma_H= 21^{+28}_{-16}$~GeV, and
$\sigma_R=0.40^{+0.62}_{-0.34}$~fb. The comparison for the optimal
parameters is shown in Table~\ref{comparexsection}.

The quality of our fit is good, but error bars are large. Yet, by
restricting again to the central values, there is good agreement
with our expectations. Indeed, by rescaling the partial width from
$\Gamma(H\to ZZ)\sim 1.6$~GeV to 1.55~GeV (for $M_H$ from 700 to
677~GeV), and fixing $\langle\Gamma_H\rangle=21$~GeV, we find a branching
ratio $B(H\to ZZ)\sim 0.073$. For the theoretical value
$\sigma^{\rm ggF} (pp\to H)\sim  1100$ fb of Ref.~\cite{yellow} (at
$M_H = 677$~GeV), this would then imply a theoretical peak cross section
$(\sigma_R)^{\rm Theor}=1100 \times 0.073 \times 0.0045 \sim 0.36$~fb,
which only differs by 10\%  from the central value
$\langle\sigma_R\rangle=0.40$~fb of our fit. Moreover, from the central
values of the fit $\langle\sigma_R\rangle= 0.40$~fb and
$\langle\gamma_H\rangle= 0.031$, we find
$\langle\sigma_R\rangle \times \langle\gamma_H\rangle\sim 0.012$~fb, again
in good agreement with our Eq.~(\ref{exp34}).

Summarising: from the two ATLAS papers Refs.~\cite{ATLAS2} and
\cite{atlasnew}, we find consistent indications of a new resonance
in our theoretical mass range $(M_H)^{\rm Theor}\sim 690\,(30)$~GeV.
By comparing with Ref.~\cite{atlasnew}, we identify more
precisely the non-resonant background $gg\to 4l$, which can
interfere with a second resonance $H$ produced via gluon-gluon
fusion. Thus, the determinations obtained with our Eq.~(\ref{sigmat})
are more accurate, from a theoretical point of view. In practice,
there is not too much difference with the previous
Refs.\cite{CFF,universe} based on the ggF-low events of
\cite{ATLAS2}. Indeed, the two mass values
$(M_H)^{\rm EXP}=677^{+30}_{-14}$~GeV vs.\ $(M_H)^{\rm EXP}=706(25)$~GeV
and the two decay widths $\Gamma_H= 21^{+28}_{-16}$~GeV vs.\
$\Gamma_H= 29\pm 20$~GeV are well consistent within their uncertainties.
Most notably, our crucial correlation Eq.~(\ref{exp34}) is well
reproduced by the central values of the fits to the two sets of
data.

\subsection{The ATLAS high-mass $\gamma\gamma$ events}

\begin{figure}[ht]
\centering
\includegraphics[trim = 0mm 0mm 0mm 0mm,clip,width=7.0cm]
{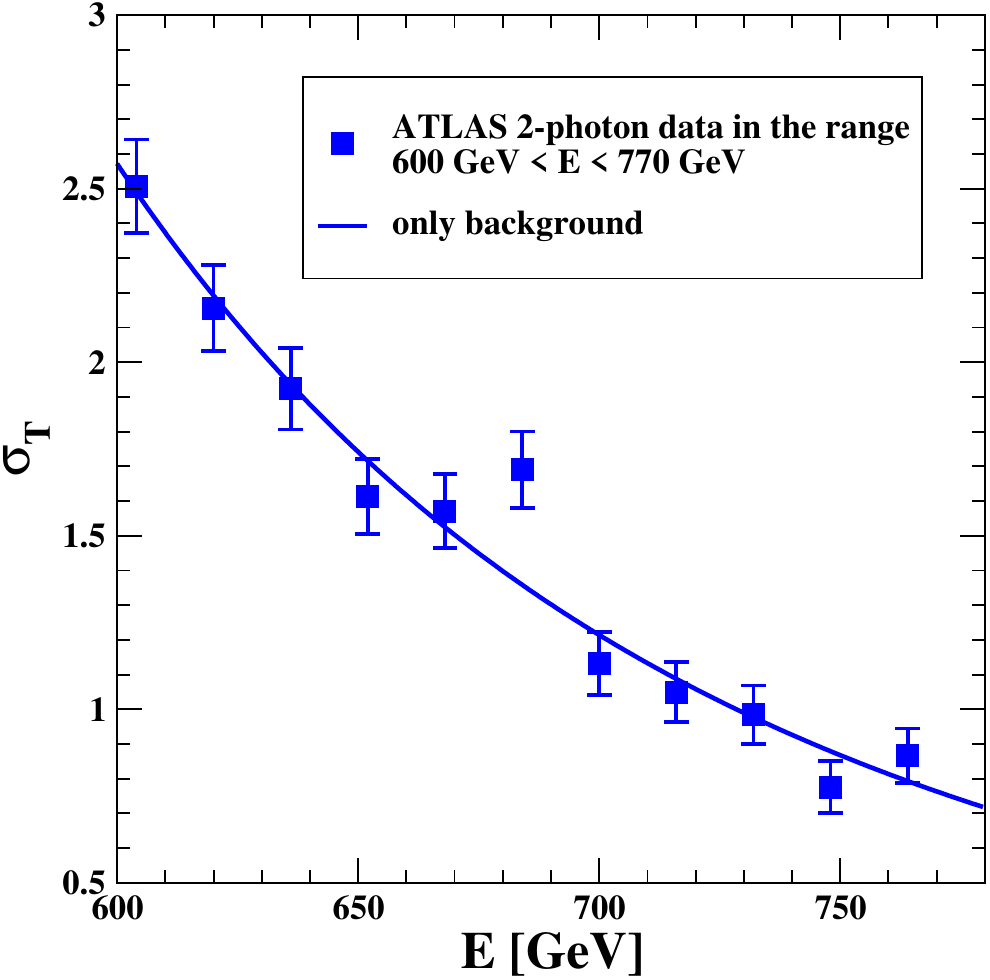}
\caption{The fit with Eq.~(\ref{sigmat}) and $\sigma_R=0$ to the ATLAS
data \cite{atlas2gamma} shown in Table~\ref{2gamma}, transformed into
cross sections in fb. The $\chi^2$ value is 14, with the
background parameters $A=1.35$~fb and $\nu=4.87$. }
\label{twogamma0}
\end{figure}
\begin{figure}[ht]
\centering
\includegraphics[trim = 0mm 0mm 0mm 0mm,clip,width=8.0cm]
{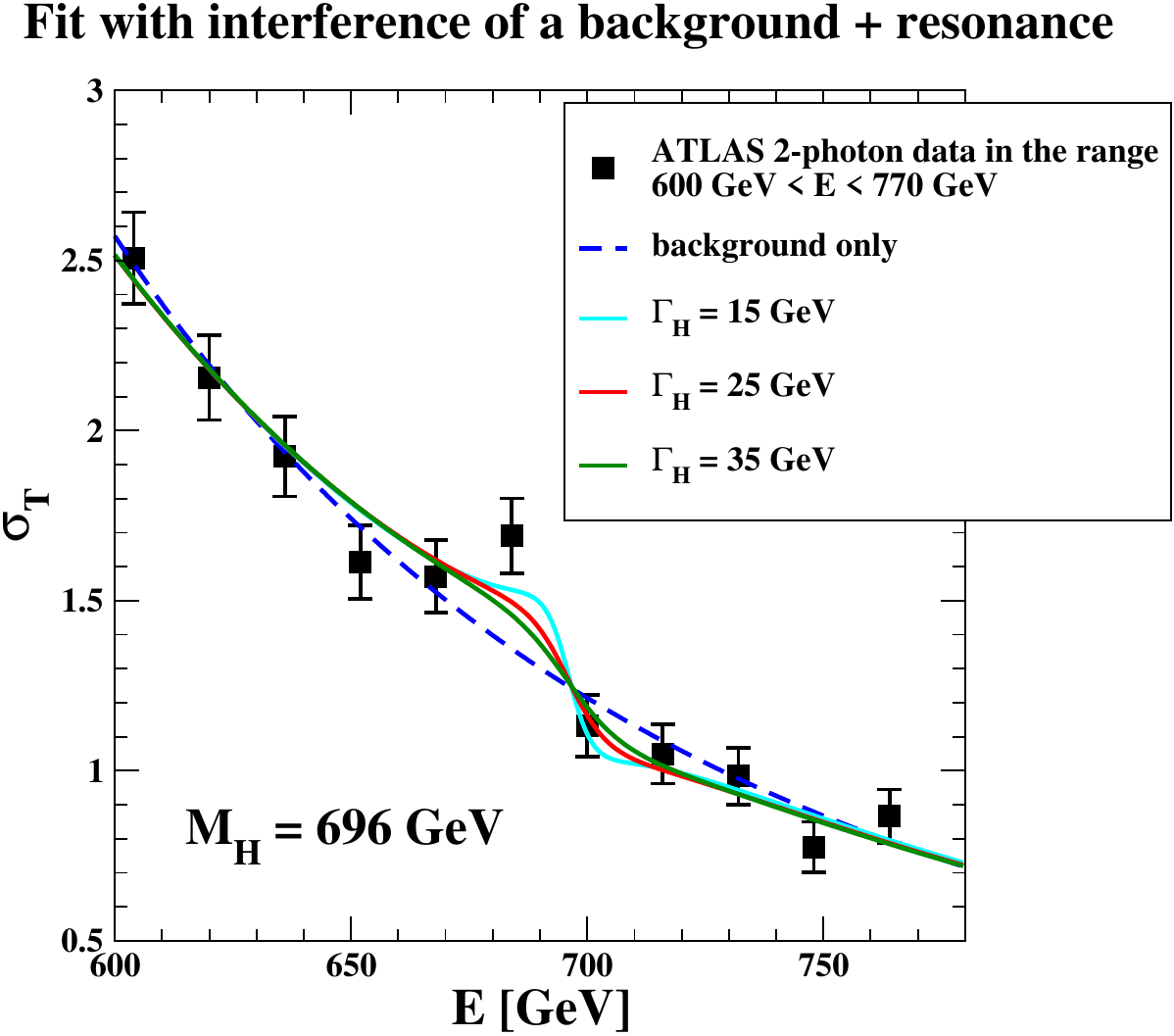}
\caption{Three fits with Eq.~(\ref{sigmat}) to the ATLAS data
\cite{atlas2gamma} in Table~\ref{2gamma},
transformed into cross sections in fb. The $\chi^2$ values are
7.5, 8.8, 10.2, for $\Gamma_H=$ 15, 25, and 35~GeV, respectively.
}\label{twogamma1}
\end{figure}
References~\cite{CFF,universe} also considered the invariant-mass
distribution of the inclusive diphoton events observed by
ATLAS\cite{atlas2gamma}; see Table \ref{2gamma}. By parametrising
the background with a power-law form
$\sigma_B(E) \sim A \times ({\rm 685~GeV}/E)^{\nu}$, a fit to the data
in Table~\ref{2gamma} gives a good description of all data points, except
for a sizable excess at 684~GeV (estimated by ATLAS to have a local
significance of more than 3$\sigma$); see Fig.~\ref{twogamma0}.
\begin{table}
\setlength{\tabcolsep}{3.5pt}.
\tbl{ The ATLAS number $N=N(\gamma\gamma)$ of events, in bins of 16~GeV
and for luminosity 139 fb$^{-1}$, for the range of invariant mass
$\mu=\mu(\gamma\gamma)=600\div 770$~GeV. These values were extracted
from Fig.~3 of Ref.~\cite{atlas2gamma}, because the relevant numbers
are not reported in the companion HEPData file.
\label{2gamma}}
{\begin{tabular}{cccccccccccc} $\mu$ & 604 & 620 & 636 & 652 & 668 &
        684 & 700 & 716 & 732 & 748 & 764 \\
 \hline $N$ & 349 & 300 & 267 & 224 & 218 & 235 &
              157 & 146 & 137 & 108 & 120 \\
\hline
\end{tabular}}
\end{table}
This isolated discrepancy shows how a (hypothetical) new resonance
might remain hidden behind a large background nearly everywhere. For
this reason, apart from the mass $M_H= 696\,(12)$~GeV,
the total decay width is determined very poorly, namely
$\Gamma_H= 15^{+18}_{-13}$~GeV, but still consistent with the other
loose determinations from the 4-lepton data. In Fig.~\ref{twogamma1} we
show three fits with the full Eq.~(\ref{sigmat}), for $\Gamma_H=$
15, 25, and 35~GeV.

\subsection{ATLAS and CMS $ (b \bar b + \gamma\gamma)$ events}
\begin{figure}[ht]
\centering
\includegraphics[trim = 0mm 0mm 0mm 0mm,clip,width=8.0cm]
{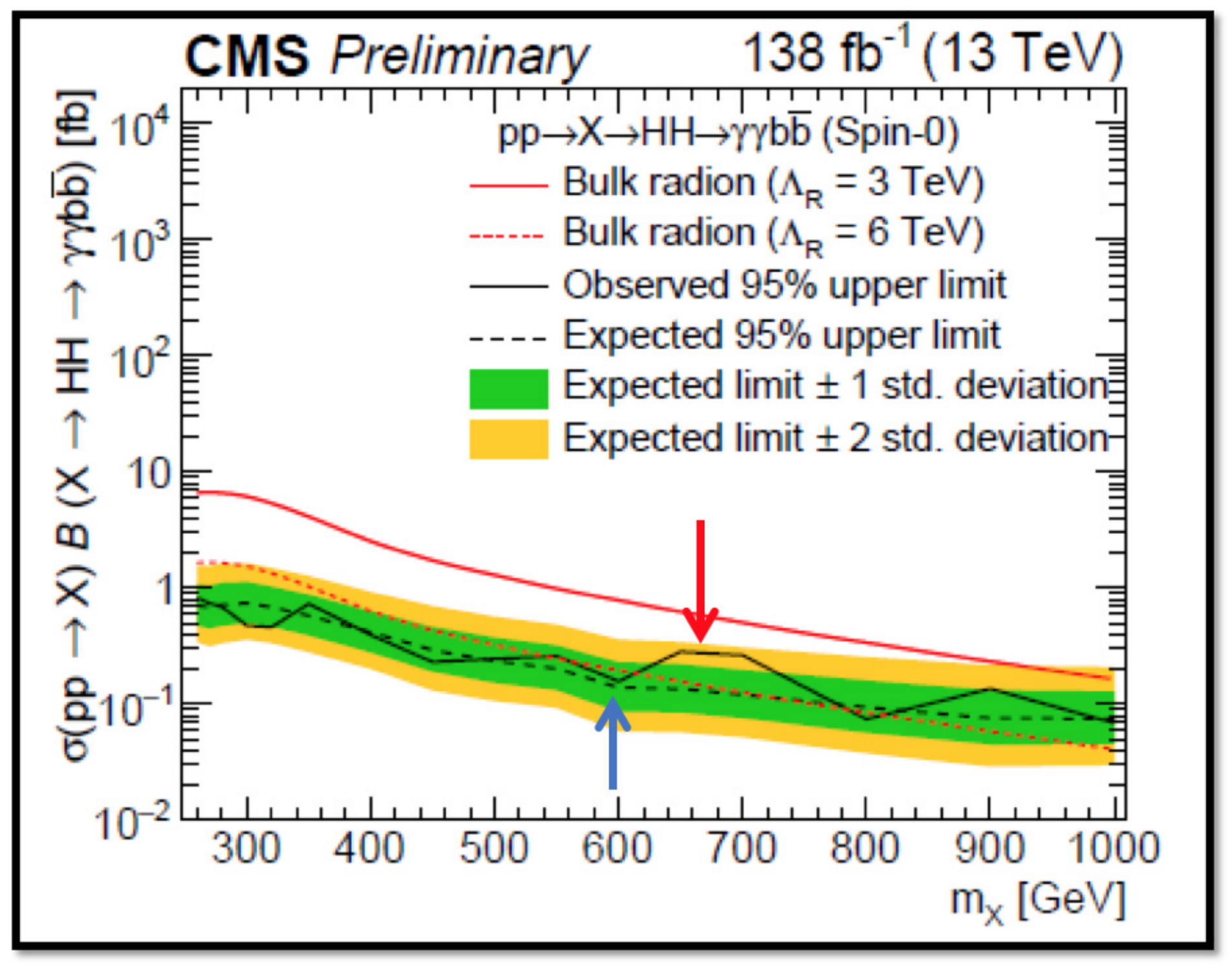}
\caption{Expected and observed 95\% upper limit for the cross section
$\sigma (pp\to X \to h(125)h(125) \to b \bar b + \gamma\gamma)$ observed
by the CMS Collaboration \cite{CMS-PAS-HIG-21-011}. }
\label{CMS_DBGG}
\end{figure}

The ATLAS and CMS Collaborations have also searched for
new resonances decaying, through two intermediate $h(125)$ scalars,
into a $b \bar b$ quark pair and a $\gamma\gamma$ pair. In
particular, in Ref.~\cite{CMS-PAS-HIG-21-011} one considered the cross
section for the full process
\BE
\sigma({\rm full})=\sigma (pp\to X\to hh\to b\bar b+\gamma\gamma) \; .
\label{CMSPAS}
\EE
For a spin-zero resonance, the 95\% upper limit $\sigma({\rm full})<0.16$~fb,
for an invariant mass of 600~GeV, was found to increase by about a factor
two, up to $ \sigma({\rm full})<0.30$~fb in a plateau $650\div 700$~GeV,
and then to decrease for larger energies; see Fig.~\ref{CMS_DBGG}.
\begin{figure}[ht]
\centering
\includegraphics[trim = 0mm 0mm 0mm 0mm,clip,width=8.0cm]
{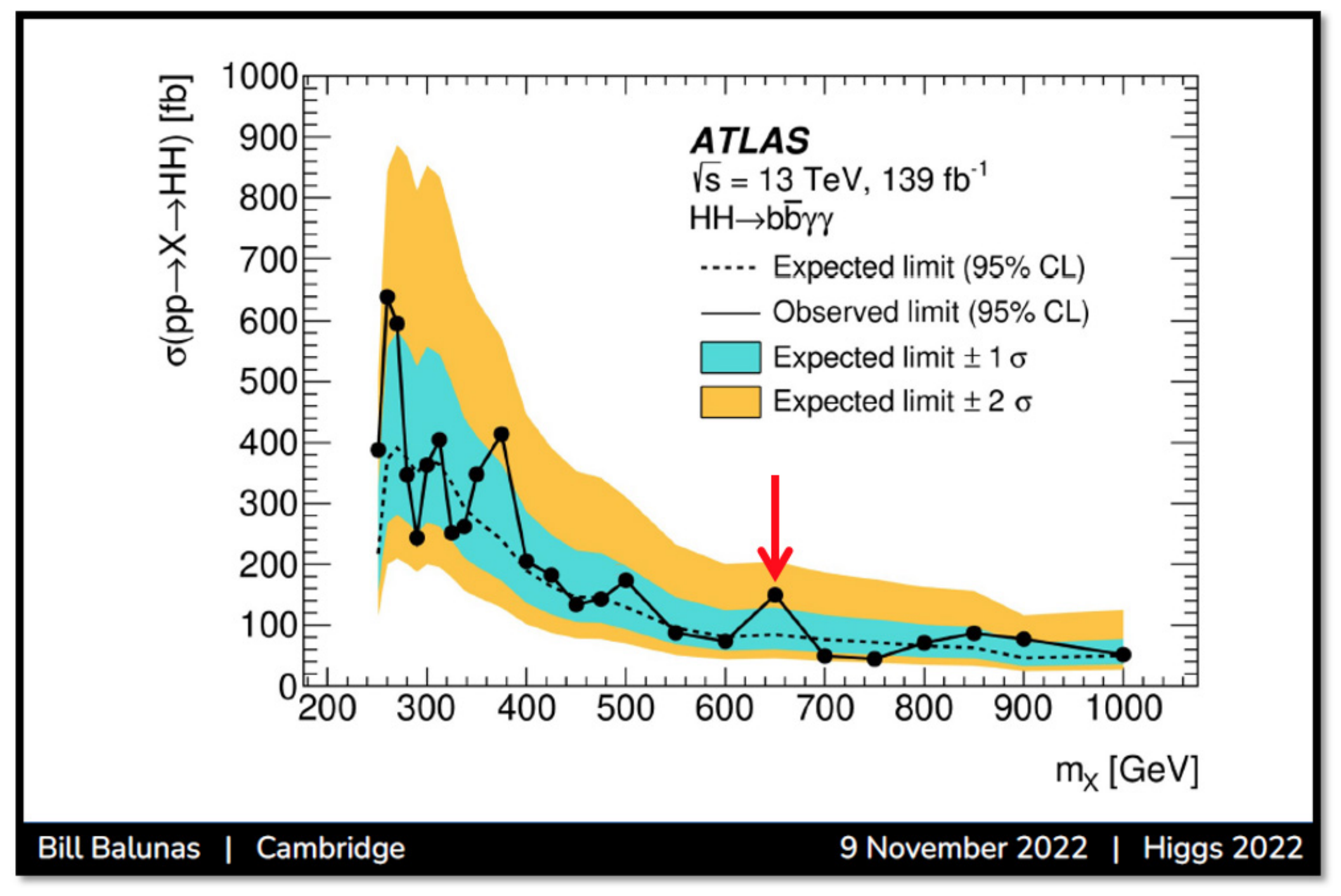}
\caption{Expected and observed 95\% upper limit for the cross
section $\sigma (pp\to X \to h(125)h(125))$ extracted by the ATLAS
Collaboration from the final state $(b \bar b + \gamma\gamma)$. The
figure is taken from the talk given by Bill Balunas at Higgs-2022
and is the same as Fig.~15 of the ATLAS paper Ref.~\cite{ATLAS_BBGG_paper}.}
\label{ATLAS_BBGG}
\end{figure}
The local statistical significance is modest, about 1.6$\sigma$, but
the relevant mass region $M_X\sim 675\,(25)$~GeV is precise and
agrees well with our prediction. The analogous ATLAS plot is
depicted in Fig.~\ref{ATLAS_BBGG} (which is the same as Fig.~15 of the
ATLAS paper Ref.~\cite{ATLAS_BBGG_paper}). Again, one finds a modest
1.2$\sigma$ excess at $650\,(25)$~GeV immediately followed by a
1.4$\sigma$ defect, which could indicate a negative above-peak
$(M^2_H - s)$ interference effect as found in the ATLAS 4-lepton
data.

\subsection{CMS-TOTEM $\gamma\gamma$ events produced in $pp$
diffractive scattering}

\begin{figure}[ht]
\includegraphics[trim = 0mm 0mm 0mm 0mm,clip,width=8.8cm]
{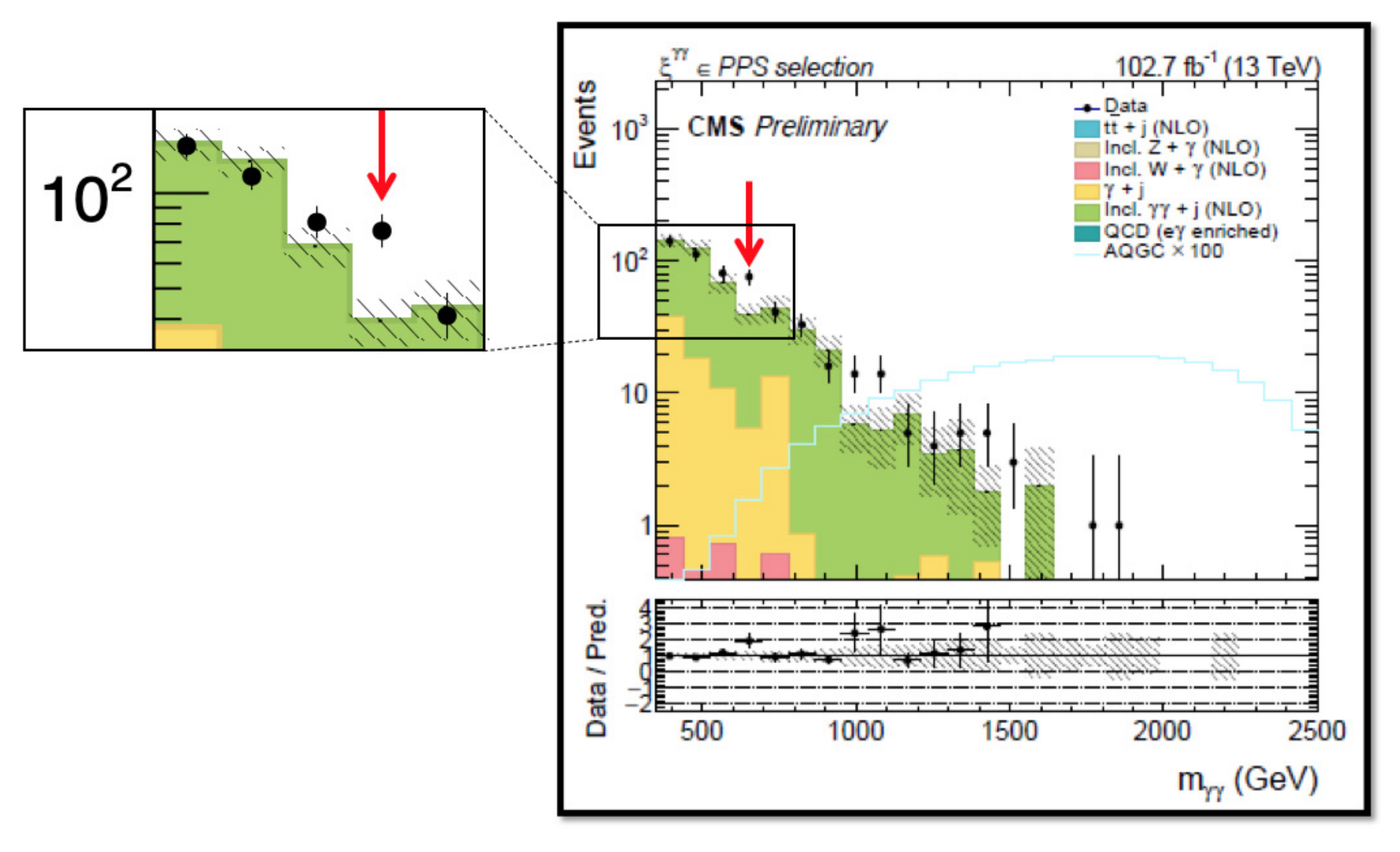}
\caption{The $\gamma\gamma$ events
produced in $pp$ diffractive scattering \cite{CMS-PAS-EXO-21-007}.
In the bin $650\,(40)$~GeV, one finds $N^{\rm EXP}\sim$ 76(9),
compared to an estimated background $N_{\rm B}\sim$ 40(9).
}\label{diffractive}
\end{figure}

The CMS and TOTEM Collaborations have also been searching for
high-mass photon pairs produced in $pp$ diffractive scattering, i.e.\
when both final protons are tagged and have large $x_F$. For our
scope, the relevant information is contained in
Fig.~\ref{diffractive}, taken from Ref.~\cite{CMS-PAS-EXO-21-007}. In the
range of invariant mass $650\,(40)$~GeV and for a statistics of
102.7~fb$^{-1}$, the observed number of $\gamma\gamma$ events is
$N^{\rm EXP}\sim 76\,(9)$, to be compared with an estimated
background $N_{\rm B}\sim 40\,(9)$. In the most conservative case,
viz.\ $N_{\rm B}= 49$, this is a local $3\sigma$ effect and is the
only statistically significant excess in the plot.

\section{Summary and conclusions}
Let us  summarise our review of LHC data:
\begin{enumerate}[i)]
\item
The ATLAS ggF-like four-lepton events \cite{ATLAS2} reported in
Table~\ref{leptontable} show a definite excess-defect sequence,
suggesting the existence of a new resonance. The same pattern is
also visible in the ATLAS cross section \cite{atlasnew}; see
Table~\ref{leptonxsection} and the $\Delta \sigma$ in
Fig.~\ref{Deltasigma}. The combined statistical deviation in the
latter analysis is at the $3\sigma$ level and indicates a resonance
mass $M_H= 677^{+30}_{-14}$~GeV.
\item
Observing the $+3\sigma$ excess at $684\,(16)$~GeV  in the
inclusive ATLAS $\gamma\gamma$ events \cite{atlas2gamma}, a fit to
these data was performed in Refs.~\cite{CFF,universe}. The resulting
mass is $M_H=696\,(12)$~GeV.
\item
An overall $+2\sigma$ effect in the ($b\bar b + \gamma\gamma)$ channel
is obtained by combining the excess of events observed by ATLAS at
$650\,(25)$~GeV and the corresponding excess observed by CMS at
$675\,(25)$~GeV.
\item
A $+3\sigma$ excess is present at $650\,(40)$~GeV in the
distribution of CMS-TOTEM $\gamma\gamma$ events produced in $pp$
diffractive scattering.
\end{enumerate}

By combining the above determinations i)-iv), the resulting estimate
$(M_H)^{\rm EXP}\sim 682\,(10)$~GeV, is in very good agreement with
our expectation $(M_H)^{\rm Theor}= 690\,(30)$~GeV. We stress
again that, when comparing with a definite prediction, one should
look for deviations from the background nearby, so that local
significance is not downgraded by the so called ``look elsewhere''
effect. Therefore, since the correlation of the above measurements
is small, the combined statistical evidence for a new resonance in
the expected mass range is far from negligible and close to (if not
above) the traditional $5\sigma$ level. We also emphasise that the
determinations i) and ii) above were obtained by fitting the
numerical data reported in Refs.~\cite{ATLAS2,atlasnew,atlas2gamma}
to the general expressions Eqs.~(\ref{sigmat}) and (\ref{NTH}). This
is very different from comparing with other Beyond-Standard-Model
scenarios (such as supersymmetry, extra-dimensions, \ldots), whose
exclusion limits assume built-in constraints (such as mass-width
and/or mass-couplings relations) that are not valid in our approach.
For this reason, we look forward to new precise data on which we can
carry out the same general analysis as with the ATLAS papers
\cite{ATLAS2,atlasnew,atlas2gamma}.

\end{document}